\documentclass[american,aip,jap,reprint]{revtex4-1}
\usepackage{mathptmx}
\usepackage[LGR,T1]{fontenc}
\usepackage[utf8]{inputenc}
\usepackage{color}
\usepackage{babel}
\usepackage{textcomp}
\usepackage{amsmath}
\usepackage{amssymb}
\usepackage{graphicx}
\usepackage[unicode=true,pdfusetitle,
 bookmarks=true,bookmarksnumbered=false,bookmarksopen=false,
 breaklinks=true,pdfborder={0 0 0},pdfborderstyle={},backref=false,colorlinks=true]
 {hyperref}
\hypersetup{
 allcolors=blue}

\makeatletter

\ProvideTextCommand{\~}{LGR}[1]{\char126#1}

\providecommand{\tabularnewline}{\\}

\usepackage{booktabs}
\usepackage{amsmath}

\makeatother

\begin{document}
\title{Acceptor levels of the carbon vacancy in 4\emph{H}-SiC: combining
Laplace deep level transient spectroscopy with density functional
modeling}
\author{Ivana Capan}
\affiliation{Division of Materials Physics, Ruđer Bošković Institute, Bijenička
54, 10 000 Zagreb, Croatia}
\author{Tomislav Brodar}
\affiliation{Division of Materials Physics, Ruđer Bošković Institute, Bijenička
54, 10 000 Zagreb, Croatia}
\author{José Coutinho}
\email{jose.coutinho@ua.pt}

\affiliation{Department of Physics and I3N, University of Aveiro, Campus Santiago,
3810-193 Aveiro, Portugal}
\author{Takeshi Ohshima}
\affiliation{Takasaki Advanced Radiation Research Institute, National Institutes
for Quantum and Radiological Science and Technology, 1233 Watanuki,
Takasaki, Gunma 370-1292, Japan}
\author{Vladimir P. Markevich}
\affiliation{School of Electrical and Electronic Engineering and Photon Science
Institute, University of Manchester, Manchester M13 9PL, United Kingdom}
\author{Anthony R. Peaker}
\affiliation{School of Electrical and Electronic Engineering and Photon Science
Institute, University of Manchester, Manchester M13 9PL, United Kingdom}
\begin{abstract}
We provide direct evidence that the broad Z$_{1/2}$ peak, commonly
observed by conventional DLTS in as-grown and at high concentrations
in radiation damaged $4H$-SiC, has two components, namely Z$_{1}$
and Z$_{2}$, with activation energies for electron emission of 0.59
and 0.67~eV, respectively. We assign these components to $\mathrm{Z}_{1/2}^{=}\rightarrow\mathrm{Z}_{1/2}^{-}+e^{-}\rightarrow\mathrm{Z}_{1/2}^{0}+2e^{-}$
transition sequences from negative-$U$ ordered acceptor levels of
carbon vacancy (V$_{\mathrm{C}}$) defects at hexagonal/pseudo-cubic
sites, respectively. By employing short filling pulses at lower temperatures,
we were able to characterize the first acceptor level of V$_{\mathrm{C}}$
on both sub-lattice sites. Activation energies for electron emission
of 0.48 and 0.41~eV were determined for $\mathrm{Z}_{1}(-/0)$ and
$\mathrm{Z}_{2}(-/0)$ transitions, respectively. Based on trap filling
kinetics and capture barrier calculations, we investigated the two-step
transitions from neutral to doubly negatively charged Z$_{1}$ and
Z$_{2}$. Positions of the first and second acceptor levels of V$_{\mathrm{C}}$
at both lattice sites, as well as $(=\!\!/0)$ occupancy levels were
derived from the analysis of the emission and capture data.
\emph{Published in Journal of Applied Physics} \textbf{\emph{124}}\emph{,
245701 (2018).} \texttt{\textbf{\textcolor{blue}{https://doi.org/10.1063/1.5063773}}}
\end{abstract}
\maketitle

\section{Introduction\label{sec:intro}}

Owing to many advantages over silicon, silicon carbide (SiC), in particularly
its $4H$ polytype, is becoming a mainstream material for the industry
of high-power electronics.\citep{Kimoto2015,She2017} Due to its wide
band gap, radiation hardness, high breakdown field and melting point,
SiC is also a promising semiconductor for the fabrication of nuclear
radiation detectors working in harsh environments, including at high
temperature and dense radiation fields.\citep{Ruddy2009,Garcia2013,Wu2017}

 SiC-based diodes for radiation detection are highly sensitive to
defects that introduce deep carrier traps,\citep{Ruddy2009} especially
to those with large capture cross section for minority carriers which
hold the actual impact signal. Point defects in SiC are mainly created
during i) semiconductor material growth, ii) device processing by
ion-implantation or iii) during operation under radiation conditions.\citep{Choyke2004}
It is therefore crucial to understand the effects of accumulated radiation
damage on the electrical properties of these devices.

In this work we investigate single and double capture/emission processes
involving a major recombination center in $4H$-SiC, namely the Z$_{1/2}$
electron trap, by combining space-charge measurements and first-principles
calculations. The Z$_{1/2}$ trap is a prominent defect in $4H$-SiC
irradiated for instance with electrons or neutrons,\citep{Nagesh1987,Storasta2004}
and can be observed by conventional deep level transient spectroscopy
(DLTS) as a conspicuous peak around room temperature.\citep{Kimoto1995,Hemmingsson1997}
It is usually present in as-grown material in concentrations in the
range $10^{12}$-$10^{13}$~cm$^{-3}$, and it is strongly anchored
to the lattice being stable up to about 1400~°C.\citep{Alfieri2005,Ayedh2015,Bathen2018}
Early DLTS experiments by Hemmingsson \emph{et~al.}\citep{Hemmingsson1998,Hemmingsson1999}
assigned Z$_{1/2}$ to the superposition of two nearly identical Z$_{1}$
and Z$_{2}$ negative-$U$ defect transitions, each located on a different
sub-lattice site. The negative-$U$ ordering of levels implies that
during the two-electron filling of the defect, the binding energy
(trap depth) of the second electron is higher than that of the first
one. Hence, during the reverse process, thermal emission of the first
electron immediately induces a second emission.

More recently, by connecting electron paramagnetic resonance (EPR)
and photo-EPR data with the DLTS results, it was possible to ascribe
Z$_{1/2}$ to transitions involving the carbon vacancy (V$_{\mathrm{C}}$)
in $4H$-SiC on distinct sub-lattice sites.\citep{Son2012} Further,
since the metastable EPR-active state was found to be the negative
charge state, it became clear that the main Z$_{1/2}$ peak had to
be connected to a $\mathrm{Z}_{1/2}^{=}\rightarrow\mathrm{Z}_{1/2}^{-}+e^{-}\rightarrow\mathrm{Z}_{1/2}^{0}+2e^{-}$
emission sequence. This is commonly labelled as Z$_{1/2}(=\!/0)$,
where the first emission is the rate-limiting step, corresponding
to the measured thermal activation energy ($\Delta E_{\mathrm{a}}$).
We also note that based on carrier concentration profiles obtained
at several temperatures using implanted/annealed samples, Z$_{1/2}$
cannot be a donor.\citep{Ayedh2014}

In Ref.~\onlinecite{Hemmingsson1998}, the reported activation energies
for electron emission were $\Delta E_{\mathrm{a}}=0.72$~eV and 0.52~eV
for Z$_{1}(=/-)$ and Z$_{1}(-/0)$, respectively, while $\Delta E_{\mathrm{a}}=0.76$~eV
and 0.45~eV for Z$_{2}(=/-)$ and Z$_{2}(-/0)$. It should be noted
that negative-$U$ defects undergo strong atomic relaxations upon
emission/capture of carriers and may show relatively high barriers
between different configurations.\citep{Watkins1984,Markevich1997}
Hence, activation energies for carrier emission often differ significantly
from the values of the thermodynamic energy levels.\citep{Peaker2018}
The latter are obtained by subtracting a capture barrier $\Delta E_{\mathrm{\sigma}}$
from $\Delta E_{\mathrm{a}}$. While it was not possible to measure
the temperature-dependence of the cross-sections (and respective barriers)
for the first electron capture, the second capture showed barriers
of $\Delta E_{\sigma}=65$~meV and 80~meV for Z$_{1}$ and Z$_{2}$,
placing the $(=\!/-)$ levels at $E_{\mathrm{c}}-0.67$~eV and $E_{\mathrm{c}}-0.71$~eV,
respectively.\citep{Hemmingsson1998}

Due to resolution limitations,\citep{Lang1974,Peaker2018} separate
emissions from Z$_{1}^{=}$ and Z$_{2}^{=}$ cannot not be resolved
by conventional DLTS. To surmount this difficulty, activation energies
and capture cross sections for Z$_{1}(=\!/-)$ and Z$_{2}(=/-)$ were
estimated by fitting the data to biexponential capacitance transients
subject to a fixed ratio between the two components (taken from the
amplitude ratio of the first acceptors).\citep{Hemmingsson1998} Hence,
the measurements of the first and second acceptors could not be carried
out independently, adding uncertainty to the measured levels.

These issues were partially addressed by some of us by means of high-resolution
Laplace deep level transient spectroscopy (L-DLTS),\citep{Dobaczewski2004}
which allowed the observation of independent emissions from Z$_{1}^{=}$
and Z$_{2}^{=}$.\citep{Capan2018} This technique had been previously
employed in the separation of an analogous set of deeper traps, labelled
$EH_{6/7}$, and attributed to donor transitions involving the V$_{\mathrm{C}}$
defect in $4H$-SiC.\citep{Alfieri2013} Laplace-DLTS was also successful
in the study of E$_{1}$/E$_{2}$ traps observed in $6H$-SiC samples.\citep{Koizumi2013}
Like Z$_{1/2}$, E$_{1}$/E$_{2}$ shows up as a prominent band in
conventional DLTS spectra of as-grown and irradiated material and
has been attributed to a carbon vacancy.\citep{Aboelfotoh1999} Notably,
from the Laplace spectra it was demonstrated that the E$_{1}$/E$_{2}$
peak had contributions from three traps, namely E$_{1}$ which showed
the highest emission rate, plus two close deeper traps, E$_{\mathrm{2L}}$
and E$_{\mathrm{2H}}$, with relatively lower and higher emission
frequencies, respectively. The three peaks were assigned to emissions
from equivalent defects located on all three sub-lattice sites of
the $6H$ polytype ($h$, $k_{1}$ and $k_{2}$).\citep{Koizumi2013}

\begin{figure}
\includegraphics[width=8.5cm]{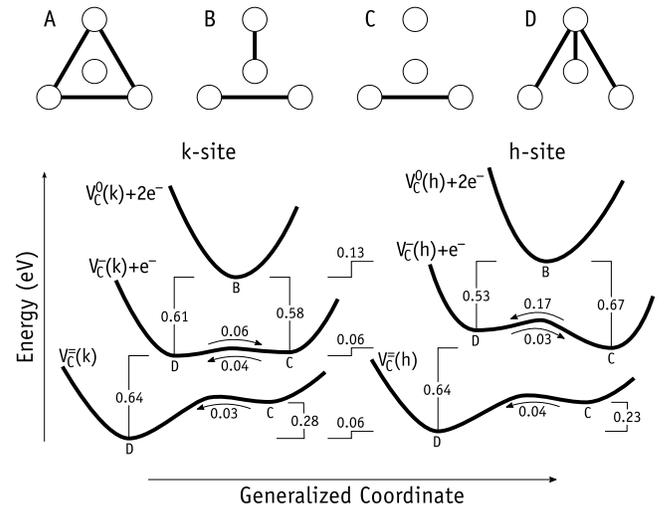}\caption{\label{fig1}(Top) Structures of the carbon vacancy in $4H$-SiC (charge
state dependent). The {[}0001{]} axis is perpendicular to the plane
of the figure. Outer circles forming a triangle are basal Si atoms,
whereas the central circle represents an axial Si. Thick lines indicate
the formation of reconstructed bonds between Si second neighbors.
(Bottom) Configuration coordinate diagram for the neutral, negative
and double negative V$_{\mathrm{C}}$ defect in $4H$-SiC located
on $k$- and $h$-sites. Transformation barriers are indicated next
to the arrows. All energies in eV.}
\end{figure}
The assignment of Z$_{1/2}$ to the carbon vacancy in $4H$-SiC has
been widely examined by first-principles modeling. While it is consensual
that V$_{\mathrm{C}}$ introduces two acceptor levels in the upper
half of the gap,\citep{Zywietz1999,Torpo2001,Bockstedte2010,Trinh2013,Hornos2011,Coutinho2017}
a clear negative-$U$ ordering of levels was obtained only when spurious
``periodic-charge'' effects were neglected and uncorrected defect
energies were used.\citep{Zywietz1999,Torpo2001,Bockstedte2010,Trinh2013}
The most recent calculations which employed hybrid density functionals,
besides not suffering from the severe underestimation of the band-gap
as displayed by previous local and semi-local calculations, considered
periodic-charge corrected energies. From these calculations, a negative-$U$
of about $-0.03$~eV was obtained for the vacancy at the $k$-site
(hereafter referred to as V$_{\mathrm{C}}(k)$), while for the $h$-site
the $U$-value was marginally positive ($+0.03$~eV).\citep{Hornos2011,Trinh2013,Coutinho2017}

The V$_{\mathrm{C}}$ defect displays several structures, depending
on the charge state and sub-lattice site.\citep{Trinh2013,Coutinho2017}
These are denoted as V$_{\mathrm{C}}(k,X)$ or V$_{\mathrm{C}}(h,X)$,
where $X\in\{\mathrm{A},\mathrm{B},\mathrm{C},\mathrm{D}\}$ is an
atomic configuration among those shown in the upper part of Figure~\ref{fig1}.
The view is along the main axial direction {[}0001{]}. The four white
circles are Si atoms, three located at basal corners and one at the
apex of a triangular pyramid. The missing carbon atom would be located
below the Si atom at the apex. Thick lines represent reconstructed
bonds formed between the Si radicals. Structure A is the fully symmetric
vacancy and it is adopted by the double-plus charge state only. In
the lower part of Figure~\ref{fig1} we also depict a calculated
configuration coordinate diagram adapted from Ref.~\onlinecite{Coutinho2017}.
Electronic transitions (and respective energies) are indicated by
the vertical separation between different minima of the potential
curves. For the $k$-site the lowest energy states are V$_{\mathrm{C}}^{=}(k,\mathrm{D})$,
V$_{\mathrm{C}}^{-}(k,\mathrm{D})$ and V$_{\mathrm{C}}^{0}(k,\mathrm{B})$,
while for the $h$-site the most stable structures are V$_{\mathrm{C}}^{=}(k,\mathrm{D})$,
V$_{\mathrm{C}}^{-}(k,\mathrm{C})$ an V$_{\mathrm{C}}^{0}(h,\mathrm{B})$.

The location of the levels from the hybrid density functional calculations
of Ref.~\onlinecite{Coutinho2017} are rather close to the Z$_{1/2}$
levels, \emph{i.e.}, V$_{\mathrm{C}}(k)$ was predicted to have levels
at $E(-\!/0)=E_{\mathrm{c}}-0.61$~eV and $E(=\!/-)=E_{\mathrm{c}}-0.64$~eV,
while V$_{\mathrm{C}}(h)$ had levels at $E(-\!/0)=E_{\mathrm{c}}-0.67$~eV
and $E(=\!/-)=E_{\mathrm{c}}-0.64$~eV. These transitions correspond
to energy differences between the most stable structures for each
charge state (see Figure~\ref{fig1}). Although the error bar of
these calculations is $\sim0.1$~eV, the prediction of a deeper $(-/0)$
transition of V$_{\mathrm{C}}(h)$ combined with the more negative
$U$-value for V$_{\mathrm{C}}(k)$ strongly suggests that Z$_{1}$
and Z$_{2}$ should be ascribed to V$_{\mathrm{C}}(h)$ and V$_{\mathrm{C}}(k)$,
respectively.\citep{Capan2018}

The goal of the present study is to resolve experimentally the electronic
transitions of Figure~\ref{fig1} by means of high-resolution L-DLTS.
The capture/emission kinetics and mechanisms, ultimately depend on
the activation energies and capture cross-sections. The later are
hard to estimate --- their calculation involves finding the electron-phonon
coupling matrix elements describing a multi-phonon emission process
(see Ref.~\onlinecite{Alkauskas2014} and references therein). Their
evaluation is outside the scope of the present work. However, in order
to get some insight into the capture/emission mechanims, we calculated
approximate values for the capture barriers of several transitions
in Figure~\ref{fig1}. The paper is organized in the following way:
In Section~\ref{sec:methods} we describe the experimental and theoretical
methodologies, then in Section~\ref{sec:experiments} we report the
conventional and Laplace DLTS data, in Section~\ref{sec:capture}
we describe the calculated capture barriers and finally we discuss
the results and draw the conclusions in Section~\ref{sec:discussion}.

\section{Experimental and theoretical methods\label{sec:methods}}

Schottky barrier diodes (SBDs) were produced from epitaxially grown
n-type $4H$-SiC layers doped with nitrogen (up to $5\times10^{14}$~cm$^{-3}$)
with $\sim25$~μm thickness.\citep{Ito2008} Schottky barriers were
formed by evaporation of nickel through a metal mask with patterned
square apertures of 1~mm~$\times$1~mm, while Ohmic contacts were
formed on the backside of the SiC substrate by nickel sintering at
950~°C in Ar atmosphere.

The quality of the SBDs was investigated by current-voltage ($I$-$V$)
and capacitance-voltage ($C$-$V$) measurements. A net doping concentration
of $4.8\times10^{14}$~cm$^{-3}$ was obtained from the $C$-$V$
measurements at 1~MHz and room temperature. Deep level defects were
analyzed by means of DLTS and high-resolution L-DLTS to determine
their respective activation energies for electron emission and capture
cross sections. The DLTS measurements were performed in the temperature
range 100-420~K at a ramp rate 3~K/min, reverse bias $V_{\mathrm{r}}=-10$~V,
pulse bias $V_{\mathrm{p}}=0$~V, pulse width $t_{\mathrm{p}}=1$~ms
and using a rate window of 50~s$^{-1}$.

For the L-DLTS measurements,\citep{Dobaczewski2004} capacitance transients
were measured with sampling rate, number of samples and number of
averaged scans in the range 5-80~kHz, 1200-9000 and 50-3000, respectively.
Reverse and pulse biases were respectively $V_{\mathrm{r}}=-5$~V
and $V_{\mathrm{p}}=0$~V. Pulse widths were $t_{\mathrm{p}}=1$~ms
and 100~ns for the $(=\negmedspace/0)$ and $(-/0)$ transitions,
respectively. The estimated error of the temperature used in the L-DLTS
measurements was less than 0.1~K.

For studying the capture kinetics, capacitance transients were measured
with different pulse widths in the range $4\times10^{-8}$-$5\times10^{-4}$~s,
while keeping the other parameters constant. In this case, reverse
voltage and pulse voltage were respectively $V_{\mathrm{r}}=-10$~V
and $V_{\mathrm{p}}=0$~V. Sampling rate, number of samples and number
of averaged scans were in the range 4-80~kHz, 4000-8000 and 600-1500,
respectively.

The Fermi level position at the temperature ranges where Z$_{1/2}(=\!/0)$
and Z$_{1/2}(-/0)$ emission peaks were observed was approximately
0.30~eV and 0.22~eV below $E_{\mathrm{c}}$, respectively. So, in
both cases the Fermi level is significantly higher than the occupancy
levels of the defects.

For the calculations we employed the \emph{Vienna Ab-initio Simulation
Package} (VASP) code,\citep{Kresse1993,Kresse1994,Kresse1996} which
implements a plane-wave based density functional method. Projector-augmented
wave (PAW) potentials were used to describe the core electrons.\citep{Bloch1994}
The PAW potentials for Si and C species were generated in the $3\mathrm{s}^{2}3\mathrm{p}^{2}$
and $2\mathrm{s}^{2}2\mathrm{p}^{2}$ valence configurations, respectively.
We employed the generalized gradient approximation to the exchange-correlation
energy as prescribed by Perdew, Burke and Ernzerhof.\citep{Perdew1996}
The Kohn-Sham states were expanded in plane-waves with a cut off energy
of $420\,\mathrm{eV}$. 

Atomistic models of V$_{\mathrm{C}}$ defects were inserted in 400-atom
$4H$-SiC supercells, obtained by replication of $5\times5\times2$
unit cells (using the theoretical lattice parameters $a=3.088$~Å
and $c=10.167$~Å). We employed a $2\times2\times1$ Monkhorst and
Pack $\mathbf{k}$-point grid to sample the Brillouin zone.\citep{Monkhorst1976}
Structural optimization was carried out by means of a conjugate gradient
method, with a convergence threshold of $5\times10^{-3}$~eV/Å for
the maximum force acting on the nuclei. The self-consistent electronic
relaxation cycles were computed with an accuracy of $10^{-8}$~eV.

\begin{figure}
\includegraphics[width=6.5cm]{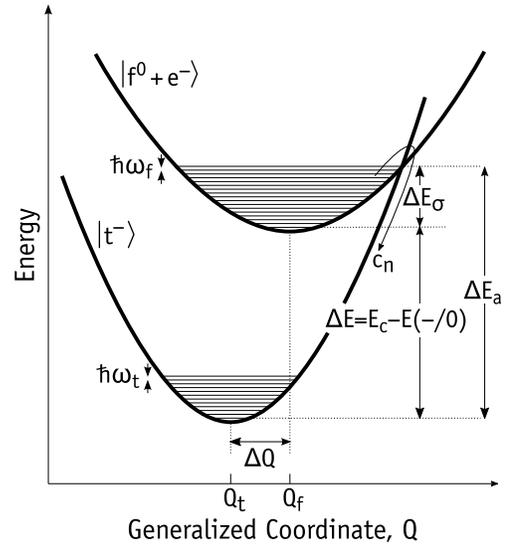}\caption{\label{fig2}Schematic configuration coordinate diagram describing
the capture of an electron. Free-electron ($|\mathrm{f}^{0}+e^{-}\rangle$)
and trapped-electron ($|\mathrm{t}^{-}\rangle$) states are shown
as parabolic curves. See text for a description of the quantities
indicated.}
\end{figure}

\subsection{Classical capture barriers}

Non-radiative capture of free carriers at deep traps often occurs
via multi-phonon emission (MPE).\citep{Huang1950,Kubo1955,Henry1977}
Within a classical harmonic picture, MPE capture can be described
by means of a configuration coordinate diagram (CCD) as depicted in
Figure~\ref{fig2}, which refers to the capture of electrons for
the sake of convenience. It represents two parabolic potential energy
curves, associated with free- ($|\mathrm{f}^{0}+e^{-}\rangle$) and
trapped-electron ($|\mathrm{t}^{-}\rangle$) states, with respective
vibrational mode frequencies $\omega_{\mathrm{f}}$ and $\omega_{\mathrm{t}}$,
and energy minima separated by $\Delta Q=Q_{\mathrm{f}}-Q_{\mathrm{t}}$
in a generalized coordinate axis. In the energy axis both states are
separated by a transition level $\Delta E=E_{\mathrm{c}}-E(-/0)$
below the conduction band bottom and they cross at $\Delta E_{\mathrm{a}}$.

MPE transitions take place close to the crossing-point of the two
curves at a rate $c_{\mathrm{n}}=\sigma_{\mathrm{n}}\bar{v}_{\mathrm{n,th}}n$,\citep{Shockley1952,Hall1952}
which encodes the capture cross section of the trap ($\sigma_{\mathrm{n}}$),
the average thermal velocity of free electrons ($\bar{v}_{\mathrm{n,th}}$)
and their concentration ($n$). Broadly speaking, the capture cross
section for a MPE transition is $\sigma_{\mathrm{n}}=A_{\mathrm{ft}}\Gamma$,
where $A_{\mathrm{ft}}$ is a purely electronic term that describes
the quantum mechanical tunneling rate between free and trapped electron
states, while $\Gamma$ is the often-called ``line-shape function''
describing the vibrational contribution to the transition rate.\citep{Huang1950}
These terms dominate $\sigma_{\mathrm{n}}$ at low and high temperatures,
respectively. In the latter case, the capture process becomes thermally
activated as $\sigma_{\mathrm{n}}\sim\exp(-\Delta E_{\sigma}/k_{\mathrm{B}}T)/\sqrt{T}$,\citep{Henry1977}
with $k_{\mathrm{B}}$ being the Boltzmann constant and $\Delta E_{\sigma}$
is a \emph{capture barrier}, \emph{i,e.}, the energy of the CCD crossing
point with respect to the potential minimum of the free-electron state.

Obtaining $\sigma_{\mathrm{n}}$ from first-principles is an involved
task (see for instance Refs.~\onlinecite{Shi2012} and \onlinecite{Alkauskas2014})
which will not be attempted here. Alternatively, we will carry out
a comparative analysis of the capture barriers for several transitions
displayed in Figure~\ref{fig1}. To achieve this we have to make
bold assumptions. The first is that the vibronic system can be described
by a single \emph{effective} mode of vibration.\citep{Stoneham1981,MakramEbeid1982}
In such a one-dimensional CCD, the relevant parameters are the effective
frequencies $\omega_{\mathrm{f}}$ and $\omega_{\mathrm{t}}$, the
modal mass $M$ and a modal vector connecting the atomic coordinates
of $N$ atoms of the free- and trapped-electron states, $\Delta\mathbf{R}=\mathbf{R}_{\mathrm{f}}-\mathbf{R}_{\mathrm{t}}=(\Delta\mathbf{r}_{1},\ldots,\Delta\mathbf{r}_{N})$.
Here $\Delta\mathbf{r}_{\alpha}=\mathbf{r}_{\mathrm{f};\alpha}-\mathbf{r}_{\mathrm{t};\alpha}$,
with $\alpha=1,\ldots,N$ and $\mathbf{r}_{\{\mathrm{f,t}\};\alpha}$
is a Cartesian coordinate of the $\alpha$-th atom.

The second assumption is that the harmonic approximation holds on
both states. We define the generalized coordinate $Q$ as,\citep{Schanovsky2011}

\begin{equation}
Q^{2}=\sum_{\alpha}m_{\alpha}\lambda^{2}|\Delta\mathbf{r}_{\alpha}|^{2},\label{eq:gencoord}
\end{equation}
 which is obtained from linear interpolation of the coordinates weighted
by atomic masses $m_{\alpha}$, where $\lambda$ is an arbitrary scalar.
The units of $Q$ are amu$^{1/2}$Å (amu - atomic mass unit). The
modal mass

\begin{equation}
M=\frac{\sum_{\alpha}m_{\alpha}\Delta r_{\alpha}^{2}}{\sum_{\alpha}\Delta r_{\alpha}^{2}},
\end{equation}
allows us to relate the atomistic distance $\Delta R$ with the separation
in the CCD as $\Delta Q=M^{1/2}\Delta R$. Assuming that the origin
of energy and coordinates is at the trapped state, the potential energy
close to $\mathbf{R}_{\mathrm{t}}$ is,

\begin{equation}
E_{\mathrm{t}}(Q)=\frac{1}{2}\omega_{\mathrm{t}}^{2}Q^{2},\label{eq:energy_trapped}
\end{equation}
while near the free-carrier state the potential energy is

\begin{equation}
E_{\mathrm{f}}(Q)=\Delta E+\frac{1}{2}\omega_{\mathrm{f}}^{2}(Q-\Delta Q)^{2},\label{eq:energy_free}
\end{equation}
where effective frequencies of vibration are obtained as $\omega_{\{\mathrm{t,f}\}}=\partial^{2}E_{\{\mathrm{t,f}\}}/\partial Q^{2}$.
Finally, the vibronic coupling can by quantified by the Huang-Rhys
factor, defined as

\begin{equation}
S_{\{\mathrm{t,f}\}}=\frac{\omega_{\{\mathrm{t,f}\}}(\Delta Q)^{2}}{2\hbar},
\end{equation}
which essentially quantifies the number of phonons emitted/created
after optical (vertical) luminescence/absorption transitions. Cases
where $S\approx0$ and $S\gg1$ correspond to weak and strong coupling,
and involve small and large defect relaxations, respectively.

\begin{figure}
\includegraphics[width=8cm]{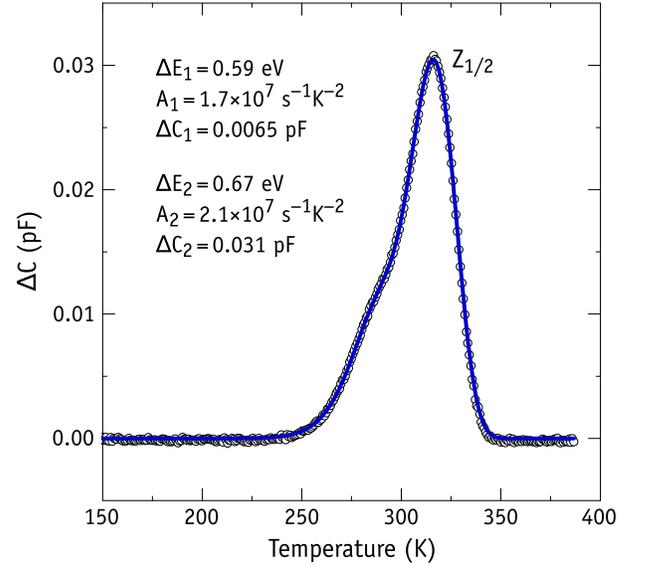}\caption{\label{fig3}Conventional DLTS spectrum (data points) obtained for
an as-grown n-type $4H$-SiC SBD. Reverse bias, pulse voltage and
width were $V_{\mathrm{r}}=-10$~V, $V_{\mathrm{p}}=0$~V,
and $t_{\mathrm{p}}=1$~ms, respectively. A rate window of 50~s$^{-1}$
was used in the measurement. The blue solid line is the simulated
DLTS spectrum with contributions from two emission signals, the parameters
of which have been determined from a least-square fitting to the experimental
data.}
\end{figure}
Equations~\ref{eq:energy_trapped} and \ref{eq:energy_free} were
fitted to first-principles total energy data $E(q,\mathbf{R})$ obtained
on a grid of coordinates $\mathbf{R}=\mathbf{R}_{\mathrm{t}}+\lambda\mathbf{R}_{\mathrm{f}}$,
between fully relaxed structures in a specific charge state $q$.
The calculated energy levels ($\Delta E$ in Eq.~\ref{eq:energy_free})
are those reported in Figure~\ref{fig1}. They were calculated using
a hybrid density functional method,\citep{Coutinho2017} which provides
accurate energy differences between defects in different charge states.
The present semi-local calculations of the harmonic potentials involve
relative energies within the same charge state. The use of non-local
functionals would not bring significant improvements.

\section{Experimental results\label{sec:experiments}}

Figure~\ref{fig3} shows a typical DLTS spectrum for as-grown $4H$-SiC
material. The broad and asymmetric peak with maximum at around 315~K
with emission rate 50~s$^{-1}$ is known as Z$_{1/2}$, and it was
assigned to $(=\!/0)$ transitions of V$_{\mathrm{C}}$ in $4H$-SiC.\citep{Son2012}
Like the E$_{1}$/E$_{2}$ peak in $6H$-SiC, the asymmetry of the
Z$_{1/2}$ peak of Figure~\ref{fig3} hints a shoulder on the low-temperature
side, suggesting the contribution of more than one defect, possibly
differing on their sub-lattice sites. The blue solid line in Figure~\ref{fig3}
is the simulated DLTS spectrum (for the measurements conditions used)
with contributions from two emission signals, the parameters of which
have been determined from least-square fitting to the experimental
data and are given in the graph. We suggest that the observed emission
signals are related to the Z$_{1}(=\!/0)$ and Z$_{2}(=\!/-)$ transitions.
Further arguments for such assignments are presented below. Concentrations
of the Z$_{1}$ and Z$_{2}$ traps in the as-grown material studied
are estimated to be about $9.5\times10^{11}$~cm$^{-3}$ and $2.1\times10^{11}$~cm$^{-3}$,
respectively.

\begin{figure}
\includegraphics[width=7.5cm]{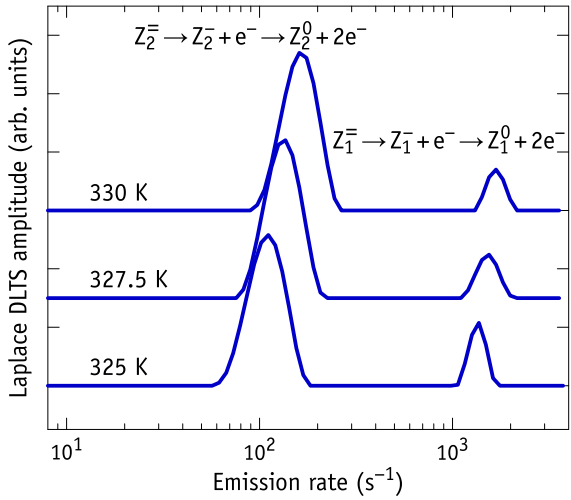}\caption{\label{fig4}L-DLTS spectra of as-grown $4H$-SiC SBD measured in
the temperature range of the Z$_{1/2}$ peak maximum. Reverse bias,
pulse voltage and width were $V_{\mathrm{r}}=-5$~V, $V_{\mathrm{p}}=0$~V,
and $t_{\mathrm{p}}=1$~ms, respectively.}
\end{figure}

Figure~\ref{fig4} shows high-resolution L-DLTS spectra measured
on the same diode for which the conventional DLTS measurements are
reported in Fig.~\ref{fig3}. The values of the measurement temperature
are in the range 325-330~K, \emph{i.e.} near the temperature of the
peak maximum of the Z$_{1/2}$ DLTS signal. The L-DLTS spectra of
Figure~\ref{fig4} clearly show that Z$_{1/2}$ consists of two close
emission components. The results confirm those reported in Ref.~\onlinecite{Capan2018},
where the high- and low-frequency peaks, namely Z$_{1}(=\!/0)$ and
Z$_{2}(=\!/0)$, were ascribed to two-electron emission signals involving
V$_{\mathrm{C}}(h)$ and V$_{\mathrm{C}}(k)$, respectively. Positions
of the emission components do not change with varying acquisition
settings (including numerical methods for the Laplace transform inversion),
and therefore it is highly unlikely that the emission signals are
related to numerical artifacts, which sometimes in the past were observed
in L-DLTS spectra.\citep{Dobaczewski2004}

Due to the negative-$U$ ordering of the Z$_{1/2}$ levels, the emission
of a second electron follows instantly after the emission of the first
one. Hence, from the L-DLTS spectra we only have access to activation
energies for the first emission. Peak amplitudes shown in Fig.~\ref{fig4}
are proportional to the change in capacitance of the space-charge,
and therefore account for both emissions. Hence, the observed two
components of Z$_{1/2}$ relate to $\mathrm{Z_{1}^{=}}\rightarrow\mathrm{Z_{1}^{-}}+e^{-}\rightarrow\mathrm{Z_{1}^{0}}+2e^{-}$
and $\mathrm{Z_{2}^{=}}\rightarrow\mathrm{Z_{2}^{-}}+e^{-}\rightarrow\mathrm{Z_{2}^{0}}+2e^{-}$
sequential transitions. We note that the labeling of the second acceptors
is consistent with that of the first acceptors in Ref.~\onlinecite{Hemmingsson1998},
where Z$_{1}(-/0)$ and Z$_{2}(-/0)$ were ascribed to the signals
with lower and higher amplitudes, respectively.

From the L-DLTS peak intensities in Figure~\ref{fig4}, we estimate
that the concentration ratio $[\mathrm{Z}_{2}^{=}]:[\mathrm{Z}{}_{1}^{=}]$
is $4.4\pm0.2$, suggesting that during growth Z$_{2}$ has higher
probability to form, most probably because it is more stable. Previous
conventional DLTS studies were not able to directly resolve these
two components. The calculated formation energies of the vacancy on
both sub-lattice sites indicate that V$_{\mathrm{C}}(k)$ is more
stable than V$_{\mathrm{C}}(h)$,\citep{Coutinho2017} supporting
the assignment of Z$_{1}$ and Z$_{2}$ to V$_{\mathrm{C}}(h)$ and
V$_{\mathrm{C}}(k)$, respectively.

\begin{figure}
\includegraphics[width=8cm]{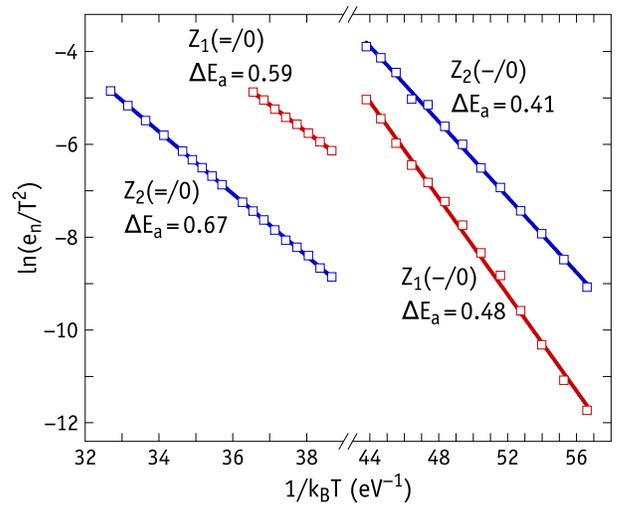}\caption{\label{fig5}Arrhenius plots of electron emission rates for Z$_{1}(=/0)$,
Z$_{2}(=/0)$, Z$_{1}(-/0)$, and Z$_{2}(-/0)$ transitions in $4H$-SiC
obtained by L-DLTS measurements. Activation energies for electron
emission are also shown for each peak. The horizontal axis is broken
and separate double from single emissions. Pre-exponential factors
and activation energies from the Arrhenius fits are reported in Table~\ref{tab1}.}
\end{figure}

From Arrhenius plots of $T^{2}$-corrected electron emission rates,
activation energies for Z$_{1}(=\!/0)$ and Z$_{2}(=\!/0)$ transitions
were determined as 0.59~eV and 0.67~eV, respectively. The Arrhenius
fits to the data are shown in Figure.~\ref{fig5}. The activation
energies compare reasonably well with calculated second acceptor levels
at $E_{\mathrm{c}}-0.64$~eV for both V$_{\mathrm{C}}(h)$ and V$_{\mathrm{C}}(k)$
defects. We should note that this comparison neglects any existing
barrier for the capture of electrons. This issue will be addressed
below.

For obtaining information about the shallower Z$_{1/2}(-/0)$ transitions
and confirm the negative-$U$ ordering of the acceptor levels, we
have applied a procedure similar to that of Ref.~\onlinecite{Koizumi2013},
which enables to freeze the negatively charged metastable configurations
in the sample. Accordingly, we fully emptied the traps by cooling
the diode from room temperature down to 220-270~K under reverse bias.
The L-DLTS spectra were then recorded by applying a short (100~ns)
filling pulse while keeping the number of scans below 50. Such conditions
ensures that the number of injected electrons is far too low to double
fill the traps, and therefore emissions from double negative defects
become small.

\begin{figure}
\includegraphics[width=7.5cm]{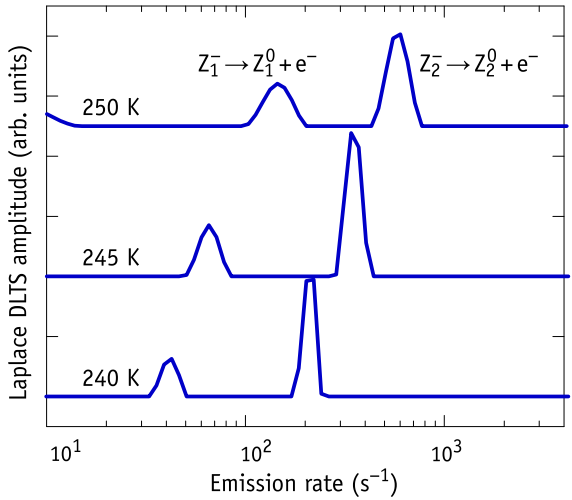}\caption{\label{fig6}L-DLTS spectra of as-grown $4H$-SiC SBD measured at
different temperatures in the range 240-250~K. Signals were obtained
by applying short ($t_{\mathrm{p}}=100$~ns) filling pulses at low
temperatures, allowing to freeze the metastable negative states in
the sample (see text for further details). Reverse  and pulse voltages
were $V_{\mathrm{r}}=-5$~V and $V_{\mathrm{p}}=0$~V, respectively.}
\end{figure}

Figure~\ref{fig6} shows the L-DLTS spectra of as grown $4H$-SiC
SBD measured at various temperatures in the range 240-250~K. The
measurements were carried out on the same SBD used to obtain the conventional
DLTS spectrum shown in Figure~\ref{fig3}. In contrast to that spectrum
and to the L-DLTS spectra recorded with the application of relatively
long (ms range) filling pulses, the use of short pulses leads to the
observation of two peaks in the L-DLTS spectra in the temperature
range 220-250~K. The two emission signals are assigned to Z$_{1}(-/0)$
and Z$_{2}(-/0)$ transitions based on their relative magnitudes and
emission rates. Interestingly, the $[\mathrm{Z}{}_{2}^{-}]:[\mathrm{Z}_{1}^{-}]$
magnitude ratio is $2.5\pm0.7$, differing from the value obtained
when longer filling pulses were applied and the traps were all double
filled. We will return to this issue in Section~\ref{sec:discussion},
where we will argue that this discrepancy could be due to kinetic
effects during the filling pulse. Activation energies for electron
emission were determined as 0.48 and 0.41~eV from Arrhenius plots
of $T^{2}$-corrected emission rates of Z$_{1}(-/0)$ and Z$_{2}(-/0)$,
respectively. These are shown on the right hand side of Figure~\ref{fig5}.

We note that the magnitudes of the Z$_{1}(-/0)$ and Z$_{2}(-/0)$
peaks decrease as the number of filling pulses increases, and eventually
disappear from the spectra after the application of a relatively large
number of filling pulses. This indicates that an increasing fraction
of double negatively charged Z$_{1/2}$ defects form and persist in
the sample for the temperature range of the measurements. Such behavior
is also a direct evidence for a negative-$U$ ordering of the acceptor
levels of Z$_{1/2}$ --- the repeated application of filling pulses
results in the capture of a second electron by lingering Z$_{1/2}^{-}$
defects and therefore in the accumulation of Z$_{1/2}^{=}$ defects.
The latter will stay in the double minus state, unless the temperature
is raised up to room temperature.

As pointed out in Section~\ref{sec:intro}, negative-$U$ defects
show at least two atomic configurations, eventually separated by an
energy barrier, and they may as well show considerable barriers for
the capture of carriers. This means that activation energies for carrier
emission, $\Delta E_{\mathrm{a}}$, may differ significantly from
carrier binding energies, $\Delta E$, which define the depth of the
trap (or a transition level) with respect to edge of the gap (see
Figure~\ref{fig2}).\citep{Peaker2018} Capture barriers and carrier
binding energies can be determined from combined emission and capture
measurements as a function of temperature.

Besides measuring the activation energies for electron emission for
all acceptor levels related to Z$_{1}$ and Z$_{2}$, we also carried
out direct capture cross section measurements. We found that electron
capture by neutral Z$_{1}$ and Z$_{2}$ traps is a very fast process.
From measurements we could not observed significant changes in magnitudes
of the Z$_{1}(-/0)$ and Z$_{2}(-/0)$ emission signals upon varying
the length of the filling pulse in the range from 40~ns to 1~μs
(in the temperature range 230-260~K). Considering the doping level
of our samples, the position of the Fermi level and the shortest length
of the filling pulse, we estimated a lower limit for the electron
capture cross section of neutral Z$_{1}$ and Z$_{2}$ being $3\times10^{-15}$~cm$^{2}$.
These results suggest the existence of a minute or even vanishing
capture barrier for both traps. So, it is likely that the depth of
the first acceptor levels of Z$_{1}$ and Z$_{2}$ with respect to
the conduction band bottom are essentially given by the activation
energies for electron emission, \emph{i.e.} they should be located
at $E_{\mathrm{c}}-0.48$~eV and $E_{\mathrm{c}}-0.41$~eV, respectively.

\begin{figure}
\includegraphics[width=8cm]{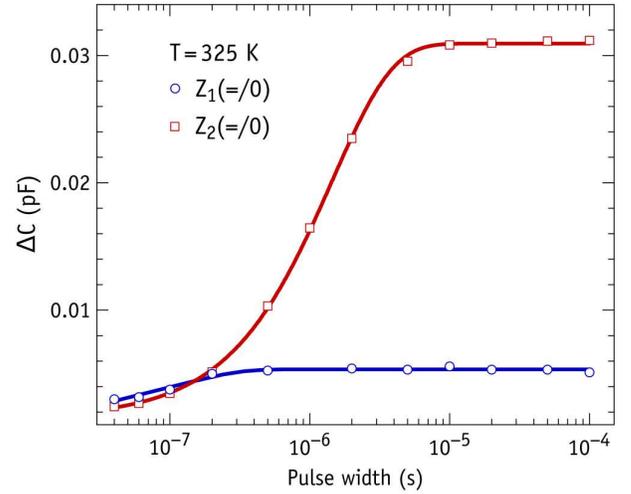}\caption{\label{fig7}Capture kinetics measured for $\mathrm{Z_{1}}(=/0)$
and $\mathrm{Z_{2}}(=/0)$ transitions by L-DLTS at $T=325$~K. Data
points are the magnitude of the L-DLTS peaks for different durations
of the filling pulse. Solid lines represent best fits of the data
to Eq.~\ref{eq:capture}}
\end{figure}
Figure~\ref{fig7} shows how the magnitude of the Z$_{1}(=/0)$ and
Z$_{2}(=/0)$ emission signals change with the length of the filling
pulse. The data were obtained by L-DLTS at $T=325$~K. For very short
pulses ($<100$~ns) the number of double filled Z$_{1/2}$ traps
is negligible, while for pulses longer than 10~$\mu\mathrm{s}$ the
signals saturate due to complete double filling. We found that the
time-dependence of the signals (measured as capacitance transients
of the diode, $\Delta C$) was satisfactorily described by a mono-exponential
law

\begin{equation}
\Delta C(t)=\Delta C_{\mathrm{max}}\left[1-\exp\left(-t/\tau\right)\right],\label{eq:capture}
\end{equation}
where $\Delta C_{\mathrm{max}}$ is the maximum amplitude of the signal
and $\tau$ is the characteristic time of the capture transient. $1/\tau$
is the defect occupancy rate, which for defects in n-type material
can be expressed as\citep{Shockley1952,Hall1952}

\begin{equation}
1/\tau=e_{\mathrm{n}}+C_{\mathrm{n}}n,\label{eq:occ-rate}
\end{equation}
where $e_{\mathrm{n}}$ is the electron emission rate, $C_{\mathrm{n}}$
is the electron capture coefficient, and $n$ is the concentration
of free electrons in the conduction band. The first and the second
terms are dominant in Eq.~\ref{eq:occ-rate} when the Fermi level
is below or above the defect occupancy level, respectively. The capture
coefficient for defects with U > 0 is expressed as

\begin{equation}
C_{n}=\sigma_{\mathrm{n}}\bar{v}_{\mathrm{n,th}},\label{eq:captcoef}
\end{equation}
where $\sigma_{\mathrm{n}}$ is the electron capture cross section,
and $\bar{v}_{\mathrm{n,th}}$ is the average thermal velocity of
free electrons. In general, the capture cross section is a temperature-dependent
quantity. For the capture process occurring via multi-phonon emission,
the capture cross section can be described by\citep{Henry1977}

\begin{equation}
\sigma_{\mathrm{n}}=\sigma_{\mathrm{n}\infty}\exp\left(-\Delta E_{\sigma}/k_{\mathrm{B}}T\right),\label{eq:barrier}
\end{equation}
where $\Delta E_{\sigma}$ is the barrier for capture, and $\sigma_{\mathrm{n}\infty}$
is the capture cross section at infinitely high temperature.

The occupancy statistics for defects with negative-$U$ properties
was considered in Ref.~\onlinecite{Markevich1997}. It was shown
that the Eq.~\ref{eq:occ-rate} is also valid for the defects with
$U<0$, however, with more complicated equations for $e_{\mathrm{n}}$
and $C_{\mathrm{n}}$. Emission of electrons by a negative-$U$ defect
with net-charge $q-2$ becomes a dominant process when the Fermi level
lies below an occupancy level $E(q\!-\!2/q)\!=\![E(q\!-\!1/q)+E(q\!-\!2/q\!-\!1)]/2$.
For $E_{\mathrm{F}}>E(q\!-\!2/q)$, capture is more effective than
emission and therefore,\citep{Markevich1997}

\[
1/\tau=C_{\mathrm{n}}^{\mathrm{eff}}n.
\]
It was found that in this case, up to four different terms can contribute
to $1/\tau(T)$, depending on the position of Fermi level with respect
to the $E(q\!-\!2/q\!-\!1)$ level and its configurations in the $q-1$
charge state.\citep{Markevich1997}

When analyzing the capture results presented in Figure~\ref{fig7}
we have taken into account the position of the Fermi level with respect
to Z$_{1/2}(=/-)$ and Z$_{1/2}(-/0)$ defect levels and the configuration
structure of V$_{\mathrm{C}}$ at $k$ and $h$ lattice sites in the
singly negatively charged states (\emph{c.f.} Figure~\ref{fig1}).
It has been concluded that in this case $1/\tau(T)$ can be expressed
as

\begin{equation}
1/\tau=\sigma_{\mathrm{n}}\bar{v}_{\mathrm{n,th}}n.\label{eq:srh}
\end{equation}

The solid curves in Figure~\ref{fig7} represent the best fits of
Eq.~\ref{eq:capture} to the Z$_{1/2}(=/0)$ capture transient data,
from which we extracted values of $\Delta C_{\mathrm{max}}$ and $\tau$.
Due to its weak magnitude, combined with the sensitivity limits of
the equipment, we could not determine the characteristic time of the
capture process for Z$_{1}(=/0)$. According to Eq.~\ref{eq:occ-rate}
a lower limit for the capture cross section of Z$_{1}(=/0)$ was estimated
to be $10^{-15}$~cm$^{2}$, corresponding to a rather small capture
barrier. Combining these findings with those above which indicate
a vanishing capture barrier for Z$_{1}(-/0)$ as well, we conclude
that the depth of the Z$_{1}(=/-)$ trap is essentially the activation
energy for electron emission from the double negative charge state,
\emph{i.e.} the level should be at about $E_{\mathrm{c}}-0.59$~eV.

\begin{figure}
\includegraphics[width=8cm]{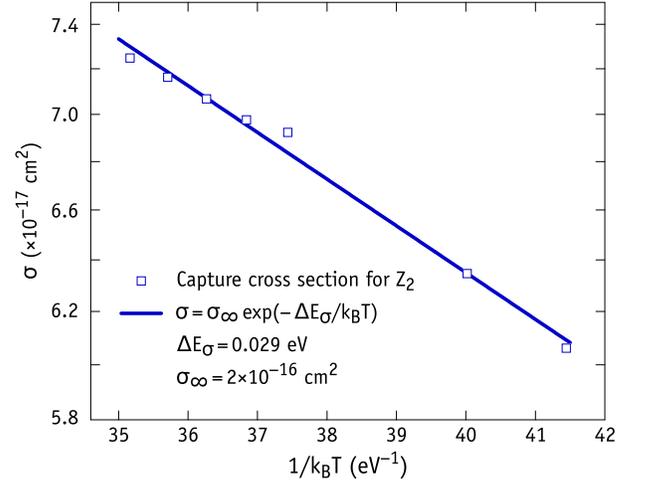}\caption{\label{fig8}Temperature dependence of the electron capture cross
section of Z$_{2}(=/0)$. The solid line represents the best fit of
the data to an Arrhenius relation describing a thermally activated
capture process. The best values obtained for the capture barrier
($\Delta E_{\sigma}$) and direct capture cross section ($\sigma_{\infty}$)
are also shown.}
\end{figure}
The capture kinetics for Z$_{2}(=/0)$ was measured in the temperature
range 300-350K. The Fermi level lies well above the metastable Z$_{2}(-/0)$
level which is estimated at 0.41~eV below the conduction band edge.
This ensures that we avoid the formation of singly negative metastable
states during the filling pulse. Considering the observed fast capture
of electrons by neutral Z$_{2}$, the thermally activated capture
kinetics of Z$_{2}(=/0)$ shown Figure~\ref{fig7} must be limited
by the capture of the second electron by singly negative Z$_{2}^{-}$
defects. These observations also suggest that the geometry of Z$_{1}$
and Z$_{2}$ defects should evolve differently along the capture sequence.
The temperature dependence of the capture cross section as derived
from the experimental data and the use of Eqs.~\ref{eq:capture}
and \ref{eq:srh} is presented in Figure~\ref{fig8}. The data can
be satisfactorily described by Eq.~\ref{eq:barrier}, with $\sigma_{\infty}=2\times10^{-16}$~cm$^{2}$
and $\Delta E_{\sigma}=0.029\pm0.005$~eV. Combining the activation
energy for electron emission from double negative Z$_{2}$ with the
electron capture barrier of single negative Z$_{2}$, we arrive at
a Z$_{2}(=/-)$ transition at $E_{\mathrm{c}}-0.64$~eV.

\begin{table}
\caption{\label{tab1}Activation energies ($\Delta E_{\mathrm{a}}$ in eV),
pre-exponential factors for the Arrhenius relation of the emission
rate ($A$ in s$^{-1}$K$^{-2}$), apparent capture cross-sections
($\sigma_{\mathrm{a}}$ in cm$^{2}$), directly measured capture cross-sections
($\sigma_{\infty}$ in cm$^{2}$) and capture barriers ($\Delta E_{\sigma}$
in eV) for Z$_{1}$ and Z$_{2}$ defects in $4H$-SiC as obtained
by L-DLTS measurements.}

\begin{ruledtabular}
\begin{tabular}{lccccc}
\noalign{\vskip\doublerulesep}
Transition & $\Delta E_{\mathrm{a}}$ & $A$ & $\sigma_{\mathrm{a}}$ & $\sigma_{\infty}$ & $\Delta E_{\sigma}$\tabularnewline[\doublerulesep]
\hline 
\noalign{\vskip\doublerulesep}
Z$_{1}(=\!/0)$ & $0.59$ & $1.1\times10^{7}$ & $2.7\times10^{-15}$ & $>10^{-15}$ & $\sim0$\tabularnewline[\doublerulesep]
\noalign{\vskip\doublerulesep}
Z$_{2}(=\!/0)$ & $0.67$ & $2.55\times10^{7}$ & $6.3\times10^{-15}$ & $2\times10^{-16}$ & $0.03$\tabularnewline[\doublerulesep]
\noalign{\vskip\doublerulesep}
Z$_{1}(-/0)$ & $0.48$ & $7.4\times10^{6}$ & $1.8\times10^{-15}$ & $>10^{-15}$ & $\sim0$\tabularnewline[\doublerulesep]
\noalign{\vskip\doublerulesep}
Z$_{2}(-/0)$ & $0.41$ & $2.15\times10^{6}$ & $5.25\times10^{-16}$ & $>10^{-15}$ & $\sim0$\tabularnewline[\doublerulesep]
\end{tabular}
\end{ruledtabular}

\end{table}

\section{Calculation of capture barriers\label{sec:capture}}

Now we describe our calculations of the capture barriers for neutral
and negatively charged carbon vacancies in $4H$-SiC. Figure~\ref{fig1}
shows that for neutral V$_{\mathrm{C}}$ the capture process departs
from V$_{\mathrm{C}}^{0}(\mathrm{B})+e^{-}$ and can in principle
arrive either at V$_{\mathrm{C}}^{-}(\mathrm{C})$ or V$_{\mathrm{C}}^{-}(\mathrm{D})$.
For now we are dropping the sub-lattice label ($k$ and $h$) in the
notation of the defect state as in this particular case the picture
is analogous for both pseudo-cubic and hexagonal vacancies. Hence,
for each sub-lattice site, we have to consider two effective coupling
modes, namely $Q_{\mathrm{C^{-}/B^{0}}}$ and $Q_{\mathrm{D^{-}/B^{0}}}$.
These were calculated by combining Eq.~\ref{eq:gencoord} along with
modal vectors $\Delta\mathbf{R}=\mathbf{R}_{\mathrm{B^{0}}}-\mathbf{R}_{\mathrm{C^{-}}}$
and $\Delta\mathbf{R}=\mathbf{R}_{\mathrm{B^{0}}}-\mathbf{R}_{\mathrm{D^{-}}}$.
They connect the end-coordinates $\mathbf{R}_{X^{q}}$ of ground-state
configurations $X$ in charge state $q$ (see Section~\ref{sec:methods}).

For electron capture by negatively charged vacancies, Figure~\ref{fig1}
shows that V$_{\mathrm{C}}^{-}(\mathrm{C})+e^{-}$ and V$_{\mathrm{C}}^{-}(\mathrm{D})+e^{-}$
have close relative energies, particularly for the $k$-site, and
are separated by small barriers. We therefore considered C and D initial
structures to estimate the capture barriers of vacancies at both $k$-
and $h$-sites. Regarding the final state, clearly V$_{\mathrm{C}}^{=}(\mathrm{D})$
is the most stable configuration on both sites, but since V$_{\mathrm{C}}^{=}(\mathrm{C})$
can easily transform to V$_{\mathrm{C}}^{=}(\mathrm{D})$ we also
considered capture routes such as $\mathrm{V}{}_{\mathrm{C}}^{-}(\mathrm{C})+e^{-}\rightarrow\mathrm{V}{}_{\mathrm{C}}^{=}(\mathrm{C})$.
Hence, for the second acceptor, and for both V$_{\mathrm{C}}(k)$
and V$_{\mathrm{C}}(h)$, we have direct-modes $Q_{\mathrm{C^{=}/C^{-}}}$
and $Q_{\mathrm{D^{=}/D^{-}}}$, as well as cross-modes $Q_{\mathrm{C^{=}/D^{-}}}$
and $Q_{\mathrm{D^{=}/C^{-}}}$.

\begin{figure}
\includegraphics[width=8.5cm]{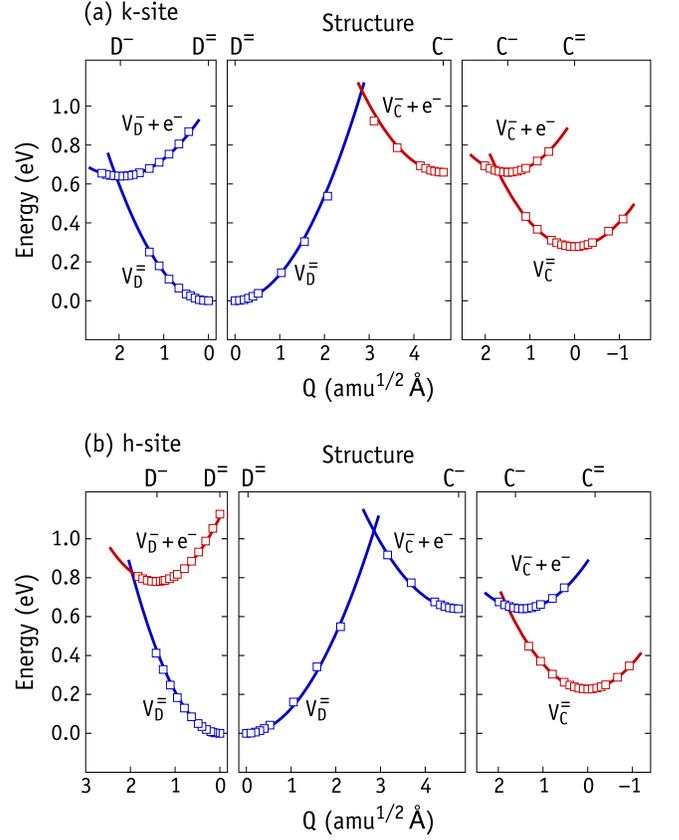}\caption{\label{fig9}Calculated configuration coordinate diagram of (a) pseudo-cubic
and (b) hexagonal carbon vacancies in $4H$-SiC illustrating capture
barriers for V$_{\mathrm{C}}^{-}$. Ground state and metastable configurations
are shown in blue and red, respectively. See text for a quantitative
description of the configuration coordinate Q. The origin of the energy
scale was set at the double minus ground state state.}
\end{figure}

The CCDs where produced by fitting Eqs.~\ref{eq:energy_trapped}
and \ref{eq:energy_free} to eleven data points within $|Q-Q_{0}|\leq0.5$~amu$^{1/2}$Å
around the minimum energy coordinate $Q_{0}$ of each state. Figures~\ref{fig9}(a)
and \ref{fig9}(b) show calculated CCDs for several $(=/-)$ transitions
involving V$_{\mathrm{C}}(k)$ and V$_{\mathrm{C}}(h)$, respectively.
Ground state and metastable configurations are represented in blue
and red colors, respectively. Energy differences between different
charge states, i.e., the electronic levels, were taken from Ref.~\onlinecite{Coutinho2017}
and are shown in Figure~\ref{fig1}. Besides the data points used
for the fittings, additional points were calculated at $|Q-Q_{0}|>0.5$~amu$^{1/2}$Å
to provide us an idea of how much the potential energy deviates from
the harmonic regime. The origin of coordinates and energy was assumed
at the trapped state, V$_{\mathrm{C}}^{=}(\mathrm{D})$.

From the fittings, effective mode frequencies in the range $\omega\sim200\textrm{-}300$~cm$^{-1}$
were obtained. These are consistent with vibrations involving weak
Si-Si reconstructed bonds edging the vacancy defects. These frequencies
should be compared to 520~cm$^{-1}$ which is the Raman frequency
involving stiffer Si-Si bonds in bulk Si. According to Stoneham,\citep{Stoneham1981}
the relative positioning of the initial state (before capture) with
respect to the final state (after capture) can be described by a coupling
ratio,

\begin{equation}
\Lambda=\frac{S}{S+p},\label{eq:coupling}
\end{equation}
where $S$ is the Huang-Rhys factor, while $p=\Delta E/\hbar\omega$
is the number of phonons spanning the zero-phonon energy. Accordingly,
three distinctive situations may occur, (i) $\Lambda\rightarrow0$
which represents the weak coupling limit, where $\Delta Q\sim0$,
\emph{i.e.} the coordinates of both states involved are nearly coincident;
(ii) $\Lambda=1/2$ representing a strong coupling case where the
potential energy curve of the final state (lower parabola) crosses
the initial state (upper parabola) at its minimum energy; (iii) $\Lambda\rightarrow1$
implies that $S\rightarrow\infty$, representing the \emph{absurd}
limit where initial and final states become infinitely distant in
the configurational space. We also note that for $\Lambda<1/2$ and
$\Lambda>1/2$ the minimum of the upper parabola falls inside and
outside the lower parabola, respectively. Examples of $\Lambda<1/2$
and $\Lambda>1/2$ arrangements are shown respectively on the left
and middle insets of Figure~\ref{fig9}(a).

In broad terms, for the first capture ($\mathrm{V}{}_{\mathrm{C}}^{0}+e^{-}\rightarrow\mathrm{V}{}_{\mathrm{C}}^{-}$),
and irrespectively of the lattice site, we found that both $Q_{\mathrm{C^{-}/B^{0}}}$
and $Q_{\mathrm{D^{-}/B^{0}}}$ effective modes lead to $\Lambda\sim0.5$,
and therefore to rather small capture barriers $\Delta E_{\sigma}=20\textrm{-}30$~meV.
Their height is in line with the weak response of the Z$_{1/2}(-/0)$
L-DLTS peaks as a function of the pulse time. Taking into account
the low transformation barriers that separate ground state structures
V$_{\mathrm{C}}^{-}(k,\mathrm{D})$ and V$_{\mathrm{C}}^{-}(h,\mathrm{C})$
from metastable V$_{\mathrm{C}}^{-}(k,\mathrm{C})$ and V$_{\mathrm{C}}^{-}(h,\mathrm{D}),$
respectively (see Figure~\ref{fig1}), the above results are unable
to decide on any of the two possible capture routes under scrutiny.

For electron capture by the negatively charged V$_{\mathrm{C}}^{-}$
defects, transitions involving direct modes ($Q_{\mathrm{D^{=}/D^{-}}}$
and $Q_{\mathrm{C^{=}/C^{-}}}$) also show $\Lambda$ values near
0.5. Consequently, these modes lead to capture barriers not higher
than 50~meV. The calculated data and fitted parabolas are shown on
the right and left insets of Figures~\ref{fig9}(a) and \ref{fig9}(b)
for V$_{\mathrm{C}}(k)$ and V$_{\mathrm{C}}(h)$, respectively.

Unlike the above modal distances ($\Delta Q_{\mathrm{D^{=}/D^{-}}}\sim\Delta Q_{\mathrm{C^{=}/C^{-}}}\sim2$~amu$^{1/2}$Å)
or those involved in the first acceptor ($\Delta Q_{\mathrm{C^{-}/B^{0}}}\sim\Delta Q_{\mathrm{D^{-}/B^{0}}}\sim2$~amu$^{1/2}$Å),
configurations C and D are remote from each other in the configurational
space ($\Delta Q_{\mathrm{C^{=}/D^{-}}}\sim\Delta Q_{\mathrm{D^{=}/C^{-}}}\gtrsim5$~amu$^{1/2}$Å).
For that reason, transitions coupled to $Q_{\mathrm{C^{=}/D^{-}}}$
and $Q_{\mathrm{D^{=}/C^{-}}}$ cross modes show a large $\Lambda\gtrsim0.8$
and large barriers are expected. For these transitions we estimated
capture barriers of at least 0.4~eV (see for example the middle insets
of Figures~\ref{fig9}(a) and \ref{fig9}(b)).

\section{Discussion and conclusions\label{sec:discussion}}

Before the concluding remarks, we provide a critical view on three
pending issues:

\emph{Error bars related to the calculated capture barriers} ---
The calculated capture barriers are in some cases of the order of
tens of meV. Considering the many approximations involved, the error
bars for calculated $\Delta E_{\sigma}$ values are probably of the
order of 0.1~eV. While it is possible to conclude that most capture
barriers are small (except those involving cross-modes), we will restrain
ourselves from drawing quantitative conclusions based on the calculated
barriers.

\emph{Calculated population ratios} --- It is known that above 1400~°C
vacancies are able to migrate.\citep{Bathen2018} During the cooling
of as-grown crystals, vacancies will \emph{freeze} as the temperature
drops below that threshold with a corresponding thermalized population
ratio. At such temperatures the material is intrinsic so that V$_{\mathrm{C}}$
defects will essentially adopt the neutral charge state. Taking into
account the calculated energy difference between V$_{\mathrm{C}}^{0}(k)$
and V$_{\mathrm{C}}^{0}(h)$ of 0.13~eV (see Figure~\ref{fig1}),
from Boltzmann statistics we obtain a $[\mathrm{Z}_{2}]:[\mathrm{Z}_{1}]\sim2.5$,
\emph{i.e.}, almost half of the experimental value reported in Section~\ref{sec:experiments}
from emissions by Z$_{1/2}^{=}$ defects. A 4.4 ratio would be obtained
at 1400~°C only if V$_{\mathrm{C}}^{0}(k)$ was more stable than
V$_{\mathrm{C}}^{0}(h)$ by 0.2~eV. Considering that (i) the population
ratio in the sample may not even reflect thermodynamic equilibrium
conditions, (ii) the error bar for the calculated formation energies
is at least 0.1~eV and (iii) configurational and vibrational entropy
should be very similar for analogous defects at $k$ and $h$ sites,
we must conclude than any quantitative account for the observed population
ratio by the current theory level is highly speculative.

\emph{Different Z$_{1}$:Z$_{2}$ amplitude ratios for first and second
acceptors} --- We must consider the possibility that the amplitude
ratio {[}Z$_{1}^{-}]$:{[}Z$_{2}^{-}${]} for the first acceptor (Figure~\ref{fig6})
could reflect a flawed {[}Z$_{1}]$:{[}Z$_{2}${]} population ratio.
First we note that both neutral Z$_{1}$ and Z$_{2}$ have almost
vanishing capture barriers. That leaves us with a capture kinetics
dominated by the quantum mechanical tunneling probability at the transition
state, which is embodied by the direct capture cross section. Table~\ref{tab1}
shows that Z$_{1}$ traps have larger capture cross sections than
Z$_{2}$. Hence, during a very short filling pulse, the formation
of Z$_{1}^{-}$ will be favored in detriment of Z$_{2}^{-}$. This
could result in a deceiving {[}Z$_{2}^{-}${]}:{[}Z$_{1}^{-}${]}$\sim\!2.5$
ratio which under-estimates the true {[}Z$_{2}${]}:{[}Z$_{1}${]}$\sim\!4.4$
concentration ratio obtained from Z$_{1/2}(=/0)$ transitions (Figure~\ref{fig4}).

If Z$_{1}^{0}$ and Z$_{2}^{0}$ are in fact V$_{\mathrm{C}}^{0}(h,\mathrm{B})$
and V$_{\mathrm{C}}^{0}(k,\mathrm{B})$ defects which differ in their
second neighboring ligands only, one wonders why do they show such
different capture cross sections? A possible reason can be found not
only in the vacancy states, but in the localization of the lower conduction
band states of $4H$-SiC. From inspection of the local density of
states (LDOS) of bulk $4H$-SiC close to the conduction band minima
and from plots of $|\psi(\mathbf{r})|^{2}$, we found that the localization
is mostly found on Si($k$)-C($k$) dimers, and nearly vanishes on
Si($h$)-C($h$) dimers. Since V$_{\mathrm{C}}^{0}(h)$ is edged by
three Si($k$) atoms and one Si($h$) radicals, we expect a larger
overlap between the acceptor states of V$_{\mathrm{C}}^{0}(h)$ and
the conduction band minimum states. On the other hand, V$_{\mathrm{C}}^{0}(k)$
has only one Si($k$) and three Si($h$) radicals, resulting in a
defect with lower capture cross section than V$_{\mathrm{C}}^{0}(h)$.
These arguments not only support the above arguments regarding the
capture kinetics, but also support the assignment of Z$_{1}$ and
Z$_{2}$ to the carbon vacancy at the $h$ and $k$ lattice sites.

Now we compare the electronic properties of Z$_{1/2}$ with those
of the carbon vacancy in $4H$-SiC. We found that both Z$_{1}$ and
Z$_{2}$ show a negative-$U$ ordering for the acceptor levels. Z$_{2}$
has the larger correlation energy ($U=-0.23$~eV) with levels at
$E(-/0)=E_{\mathrm{c}}-0.41$~eV and $E(-/0)=E_{\mathrm{c}}-0.64$~eV,
while Z$_{1}$ has levels separated by only $U=-0.11$~eV and they
are located at $E(-/0)=E_{\mathrm{c}}-0.48$~eV and $E(-/0)=E_{\mathrm{c}}-0.59$~eV,
both lying right between Z$_{2}(-/0)$ and Z$_{2}(=/-)$ . In fact,
both Z$_{1}(=\!/0)$ and Z$_{2}(=\!/0)$ occupancy (thermodynamic)
levels coincide at $E_{\mathrm{c}}-0.53$~eV, \emph{i.e.} at mid-way
between their respective metastable acceptors. These results agree
well with the most recent calculations,\citep{Hornos2011,Coutinho2017}
where $(=/0)$ occupancy levels for V$_{\mathrm{C}}(k)$ and V$_{\mathrm{C}}(h)$
were estimated at $E_{\mathrm{c}}-0.63$~eV and $E_{\mathrm{c}}-0.65$~eV,
but more significantly V$_{\mathrm{C}}(k)$ showed a negative $U=-0.03$~eV,
while V$_{\mathrm{C}}(h)$ had a marginally positive $U=+0.03$~eV,
supporting direct connections between Z$_{1/2}$ and V$_{\mathrm{C}}(h/k)$,
respectively.

Before discussing the mechanisms for electron capture it is important
to note that EPR confirms that negatively charged V$_{\mathrm{C}}^{-}(k)$
and V$_{\mathrm{C}}^{-}(h)$ defects show D and C ground state structures,
respectively.\citep{Trinh2013,Coutinho2017} Alternative V$_{\mathrm{C}}^{-}(k,\mathrm{C})$
and V$_{\mathrm{C}}^{-}(h,\mathrm{D})$ configurations are metastable
and can easily be converted to the ground states by surmounting energy
barriers of the order of 0.1~eV and lower (see Figure~\ref{fig1}).

The measurements indicate that the capture barriers for $\mathrm{V_{C}^{0}}+e^{-}\rightarrow\mathrm{V_{C}^{-}}$
are very small (or even vanishing). The calculated capture barriers
were also found to be very small for transitions involving ground
state modes $Q_{\mathrm{D^{-}/B^{0}}}$ and $Q_{\mathrm{C^{-}/B^{0}}}$
for V$_{\mathrm{C}}(k)$ and V$_{\mathrm{C}}(h)$, respectively, but
they suggest as well that transitions through intermediate metastable
states V$_{\mathrm{C}}^{-}(k,\mathrm{C})$ and V$_{\mathrm{C}}^{-}(h,\mathrm{D})$
have small capture barriers. From those configurations, a final conversion
to ground state structures is only limited by very small transformation
barriers. Hence, although it is reasonable to assume mechanisms involving
direct transitions between ground states,

\begin{eqnarray*}
\mathrm{Z}_{1}(-/0) & : & \mathrm{V_{C}^{0}}(h,\mathrm{B})+e^{-}\xrightarrow{\Delta E_{\sigma}\sim0}\mathrm{V_{C}^{-}}(h,\mathrm{C})\\
\mathrm{Z}_{2}(-/0) & : & \mathrm{V_{C}^{0}}(k,\mathrm{B})+e^{-}\xrightarrow{\Delta E_{\sigma}\sim0}\mathrm{V_{C}^{-}}(k,\mathrm{D}),
\end{eqnarray*}
we actually cannot rule out the involvement of metastable states.

Regarding the capture of a second electron, we find that Z$_{1}^{-}$
and Z$_{2}^{-}$ behave differently, with the former essentially showing
a vanishing capture barrier, while for Z$_{2}^{-}$ we could obtain
a small but measurable barrier of $\Delta E_{\sigma}=0.03$~eV. The
calculations also anticipate a different mechanism for the second
capture by the vacancy at the $k$- and $h$-sites. While in Figure~\ref{fig9}(a)
the capture by V$_{\mathrm{C}}^{-}(k)$ involving ground state structures
$Q_{\mathrm{D^{=}/D^{-}}}$ (blue lines) show a very small capture
barrier, the analogous transition involving ground state structures
for V$_{\mathrm{C}}^{-}(h)$ is shown in the middle inset of Figure~\ref{fig9}(a),
and clearly results in a large barrier. Alternatively, we suggest
that the transition takes place, firstly via electron capture coupled
to $Q_{\mathrm{C^{=}/C^{-}}}$, which shows a minute barrier (right
inset of Figure~\ref{fig9}), quickly followed the transformation
towards the ground state V$_{\mathrm{C}}^{=}(h,\mathrm{D})$ over
a barrier which was calculated to be as low as 0.04~eV (see left
side of Figure~\ref{fig1}). Hence, for the second capture we find,

\begin{eqnarray*}
\mathrm{Z}_{1}(=/-) & : & \mathrm{V_{C}^{-}}(h,\mathrm{C})+e^{-}\xrightarrow{\Delta E_{\sigma}\sim0}\mathrm{V_{C}^{=}}(h,\mathrm{C})\rightarrow\mathrm{V_{C}^{=}}(h,\mathrm{D})\\
\mathrm{Z}_{2}(=/-) & : & \mathrm{V_{C}^{-}}(k,\mathrm{D})+e^{-}\xrightarrow{\Delta E_{\sigma}=0.03}\mathrm{V_{C}^{=}}(k,\mathrm{D}).
\end{eqnarray*}

To conclude, we presented a joint experimental and theoretical investigation
of the electronic properties of Z$_{1/2}$ traps in $4H$-SiC. The
study addressed the location of individual $(-/0)$ and $(=/-)$ transitions
in the band gap, as well as the capture and emission dynamics involving
these traps. The experiments were carried out by conventional and
high-resolution L-DLTS, whereas the calculations employed a plane-wave
based density functional theory method using a semi-local approximation
to the exchange-correlation energy. We were able to confirm the connection
between the levels of Z$_{1}$ and Z$_{2}$ with those of the carbon
vacancy at the hexagonal and pseudo-cubic sites of the lattice, respectively.
We also report direct capture cross section measurements for the levels.
These show minute (or vanishing) capture barriers, confirming the
calculated strong coupling between initial and final states involved
in the transitions. Based on the calculated capture and transformation
barriers, detailed mechanisms were proposed for the first and second
electron capture.

\section*{Acknowledgements}

This work is supported by the NATO SPS programme, project number 985215.
J.C. thanks the Fundação para a Ciência e a Tecnologia (FCT) for support
under project UID/CTM/50025/2013, co-funded by FEDER funds through
the COMPETE 2020 Program. I.C. acknowledges financial support from the European Union through the Regional Development Fund for the “Center of Excellence for Advanced Materials and Sensing Devices” (Grant No. KK.01.1.1.01.0001), the European Union's Horizon 2020 Research and Innovation Programme under grant agreement No. 669014, and the European Union through the European Regional Development Fund -- The Competitiveness and Cohesion Operational Programme (Grant No. KK.01.1.1.06). The work in Manchester has been funded by the UK EPSRC under contract EP/P015581/1.

\bibliographystyle{aipnum4-1}

%

\end{document}